# Large Orbital to Charge Conversion in Weak Spin Orbit Coupling Element Zr via Spin Orbital Pumping and Spin Orbital Seebeck Effect

Nakul Kumar[1], Nikita Sharma[1], Soumyarup Hait[1,$], Lalit Pandey[1,#], Nanhe Kumar Gupta[1,†], Nidhi Shukla[1], Shubhashish Pati[1], Abhay Pandey[1], Mitali[1], Sujeet Chaudhary[1,*]

[1]*Thin Film Laboratory, Department of Physics, Indian Institute of Technology Delhi, New Delhi 110016, India*

[$]*Current address: School of Physics and Astronomy, University of Leeds, Leeds LS2 9JT, United Kingdom*

[#]*Current address: Department of Microtechnology and Nanoscience, Chalmers University of Technology, SE-41296 Göthenburg, Sweden*

[†]*Current address: Centre for Magnetic and Spintronic Materials, National Institute for Materials Science 1-2-1 Sengen, Tsukuba, Ibaraki 305-0047, Japan*

*Corresponding author: sujeetc@physics.iitd.ac.in

## Abstract

The generation of spin-orbital currents is crucial for advancing energy-efficient spintronic devices. Here, the intricate process involved in the generation and conversion of spin and orbital to charge currents in Zr(t=2, 3, 4.5, 6, &10nm)/Co$_{60}$Fe$_{20}$B$_{20}$(CFB), Zr/Pt/CFB, and Zr/Pt/CFB/Pt heterostructures are investigated using spin-orbital pumping ferromagnetic resonance and longitudinal spin-orbital Seebeck effect measurements. The moderate spin-orbit coupling (SOC) in the CFB layer facilitates the simultaneous generation of spin and orbital currents, which are transferred into adjacent Zr and Pt layers. Different spin-orbital to charge current contributions, namely, Inverse spin Hall effect (ISHE), Inverse orbital Hall effect (IOHE), and Inverse orbital Rashba-Edelstein effect (IOREE) are analyzed. Notably, introducing a single Pt layer increases the spin-orbital to charge current conversion via combined effects: ISHE in Pt, IOREE in Zr/Pt interface. An enhanced *effective* spin-orbital Hall angle ($\theta_{eff}$) of $0.120 \pm 0.004$ is observed for Zr/Pt/CFB, compared to that of $0.065 \pm 0.002$ for the Zr/CFB, and $0.077 \pm 0.003$ for the Zr/Pt/CFB/Pt heterostructures. These findings provide new insights into orbital-moment dependent phenomena and offer promising avenues for developing advanced spintronic devices exploiting both spin and orbital degrees of freedom, even in materials with lower SOC.

**Introduction**

Over the past two decades, research community have harnessed advantage of both electron's spin degree of freedom and its charge for the advancement in microelectronic computing applications. For the development in spin based microelectronic, materials with high spin-orbit coupling (SOC) are required, which can facilitate the large interconversion of charge to spin (C2S) and spin to charge (S2C) by spin Hall effect (SHE) and inverse spin Hall effect (ISHE), respectively[1,2]. These phenomena are generally observed in bulk bands of high SOC material such as Platinum (Pt)[2–4], Tungsten (W)[5–7], and Tantalum (Ta) [7–9], *etc*. Another interconnected crucial effect that originates from the lack of symmetry at the surface or interface of such materials, is known as the Rashba Edelstein effect (REE) or inverse Rashba-Edelstein effect (IREE)[10–12], efficiently drives the S2C and C2S. These effects are majorly observed in two-dimensional (2D) electronic systems, topological insulators such as $Bi_2Te_3$, $Sb_2Te_3$, $Bi_2Se_3$, Ag/Bi interface, *etc*.[11,13–18]. Recently, the orbital angular momentum (OAM), an intrinsic property of element, has taken a serious attention to the research community as it can generate orbital current from the charge current or vice versa, within the weak SOC materials such as Cu, Ti, Zr, Cr, *etc*., and such phenomenon are called as orbital Hall effect (OHE) and inverse orbital Hall effect (IOHE)[19–23]. Akin to spin phenomena, the orbital momentum generated at the surface or interfaces can also facilitate S2C or C2S phenomena via Orbital Rashba-Edelstein effect (OREE) or inverse orbital Rashba-Edelstein effect (or IOREE)[24–26].

The experimental investigation of the generation and detection of spin currents is not straight forward; a few specialized techniques such as spin pumping ferromagnetic resonance (SP-FMR)[1,8], spin-torque ferromagnetic resonance (ST-FMR)[27], and longitudinal spin Seebeck effect (LSSE)[28–32] are employed to closely track the propagation of the spin current. In the SP-FMR technique, the spin current is generated by coherent precession of the magnetic moments of a ferromagnets (FM) placed under a magnetic field and electromagnetic source with frequency tuned to Lamour precession frequency[33,34]. The ST-FMR, however, uses the spin-polarised current to excite the ferromagnetic resonance condition [5]. On the other hand, in the LSSE, the spin current is generated through a thermally induced magnon wave in the FM layer when subjected to a thermal gradient in presence of magnetic field [35]. This thermally generated spin current in the FM can intertwining with the orbital angular momentum of FM and generate both spin and orbital current and this phenomenon is defined as longitudinal spin orbital Seebeck effect (LSOSE) [19,30,31,36]. Recent studies have revealed that SP-FMR phenomena in a FM having high SOC can lead to the intertwining of the generated spin current with the orbital angular momentum, leading to a mixed spin-orbital current $J_{l,S}$ defining as spin orbital pumping FMR (SOP-FMR)[22]. Directly detecting such orbital currents are significantly challenging, so a thin layer of low SOC materials such as Ti, $CuO_x$, Zr, etc. is often used to convert orbital current into charge current, which can easily be detected by various methods such as ISHE or IOHE[24,37].

Here, we have thoroughly investigated the intricate role of spin and orbital currents to generate charge current in a non-magnetic lighter element Zr which has weak SOC [24]. For this, couple of different heterostructures comprising of thin Zr film(s) as non-magnetic weak SOC layer and $C_{60}F_{20}B_{20}$ (CFB) as the FM layer are fabricated to study the related phenomena such as IOREE, ISHE, and IOHE, etc. The observed findings reveal that Zr exhibits higher orbital Hall conductivity $(\sigma_{OH}) \sim 5300 \frac{\hbar}{e}$ whereas the spin hall conductivity $(\sigma_{SH})$ is comparatively order of magnitude smaller, i.e., $\sigma_{SH} \sim -170 \frac{\hbar}{e}$ [20,24,38]. Moreover, from the SP-FMR and LSSE measurements it is observed that the magnitude of charge current generated from inverse spin- and orbital Hall-effects is independent of the thickness of Zr layer when varied in 2-10 nm range. These results conclude the dominancy of the conversion of orbital to charge current occurring via the presence of the Rashba interface or IOREE. Our results pave the way towards the goal of understanding the orbital-dependent phenomena, which can help in functionalizing of the future spintronic devices.

## Experimental section

## Sample Growth and Characterization

A series of seven Zr/CFB heterostructure samples, and two samples of Zr, CFB, and Pt heterostructure and one with CFB and $CuO_x$ heterostructure (as tabulated in Table I) were fabricated at room temperature (RT) on $Al_2O_3$ (0001) substrates using a direct current (DC) magnetron sputtering system (M/s Excel Instruments, India make) with a base pressure of $8\times10^{-8}$ Torr. The sputtering targets of $Co_{60}Fe_{20}B_{20}$ (99.95%), Zirconium (Zr) (99.95%), Gold (99.95%), and Platinum (Pt) (99.95%), each having two inches diameter were used for the deposition. During the growth of heterostructure by sputtering, the chamber was maintained at a working pressure of $2.5\times10^{-3}$ Torr by bleeding 8 sccm of Ar gas to strike *Ar* plasma, while simultaneously pumping the chamber using a combination of cryo-pump, turbo molecular pump. The Zr and CFB were sputtered by applying 60W and 80W DC-power to the respective targets to achieve the required thickness of each layer by controlling the deposition time using a pre-determined deposition rate for each target material. The interlayer of Pt was deposited at the same working condition as mentioned above with 20W of DC power applied to Pt target.

**Table-1.** Sample names and the details of the individual layer material and their respective thicknesses. Whereas, the CFB layer thickness is fixed at 20 nm, the Zr layer thickness (t) varies from 2 to 10 nm. In the sample nomenclature, whereas the letter symbols Z, C and P, , represent the Zr, CFB and Pt films, respectively, the number next to the letter represents the thickness of that particular layer.  In one of the stacks, a 3nm thin layer of $CuO_x$ is also sputtered on CFB.

| Sr. No. | Stack | Sample Name |
|---|---|---|
| 1 | $Al_2O_3$/Zr(2±0.2nm)/$Co_{60}Fe_{20}B_{20}$(20±0.5nm) | Z2C |
| 2 | $Al_2O_3$/Zr(3±0.2nm)/$Co_{60}Fe_{20}B_{20}$(20±0.5nm) | Z3C |
| 3 | $Al_2O_3$/Zr(4.5±0.4nm)/$Co_{60}Fe_{20}B_{20}$(20±0.5nm) | Z4.5C |
| 4 | $Al_2O_3$/Zr(6±0.4nm)/$Co_{60}Fe_{20}B_{20}$(20±0.5nm) | Z6C |
| 5 | $Al_2O_3$/Zr(10±0.5nm)/$Co_{60}Fe_{20}B_{20}$(20±0.5nm) | Z10C |
| 6 | $Al_2O_3$/$Co_{60}Fe_{20}B_{20}$(20±0.5nm)/Zr(4.5±0.4nm)/$Co_{60}Fe_{20}B_{20}$(20±0.5nm) | CZ4.5C |
| 7 | $Al_2O_3$/Zr(4.5±0.4nm)/$Co_{60}Fe_{20}B_{20}$(20±0.5nm)/Zr(4.5±0.4nm) | Z4.5CZ4.5 |
| 8 | $Al_2O_3$/Zr(4.5±0.4nm)/Pt(2±0.2nm)/$Co_{60}Fe_{20}B_{20}$(20±0.5nm) | Z4.5P2C |
| 9 | $Al_2O_3$/Zr(4.5±0.4nm)/Pt(2±0.2nm)/$Co_{60}Fe_{20}B_{20}$(20±0.5nm)/Pt(2±0.2nm) | Z4.5P2CP2 |
| 10 | $Al_2O_3$/$Co_{60}Fe_{20}B_{20}$(20±0.5nm)/$CuO_x$(3±0.2nm) | C-$CuO_x$ |

A novel approach was used to protect the Zr thin films from oxidation and to create direct electrical contacts (comprising of Au pads) for subsequent voltage measurements. The substrates were cut into 10 × 4 mm dimensions, and a 50 nm thick gold film was deposited on the adjacent edges with a cover in the middle, as illustrated in figure 1(a). After removing the cover, gold pads remained on both sides. Another cover slip was then placed, to cover over only the two-thirds of each pad from the edge (see figure 1(a)), leaving one-third Au-pad surface exposed. The Zr thin film was deposited, followed by a CFB thin film over the Zr layer or any other layer as desired for any stack (Table-1). Finally, the cover slips were removed to reveal the final heterostructure shown in figure 1(a). In parallel, a bare substrate was placed with the gold patterned substrate to characterise the sample properties such as X-ray reflectivity (XRR) measurement (See Supplementary file S1) and time-of-flight secondary ion mass spectroscopy (TOF-SIMS). The XRR was performed to optimise the thicknesses of individual layers and demonstrate the interface quality of samples shown in supplementary file. It's worth mentioning that the interface between the non-magnetic (NM) and the ferromagnetic layer is a very crucial aspect for the spintronics device performance. The investigation of the interface quality using the time-of-flight secondary ion mass spectroscopy (TOF-SIMS) measurement was performed in the Z4.5C and Z4.5P2CP2 samples. The resulting TOF-SIMS depth profiles (presented in figure 1 b, c) clearly demonstrate the presence of a high-quality interfaces between Zr and CFB layers.

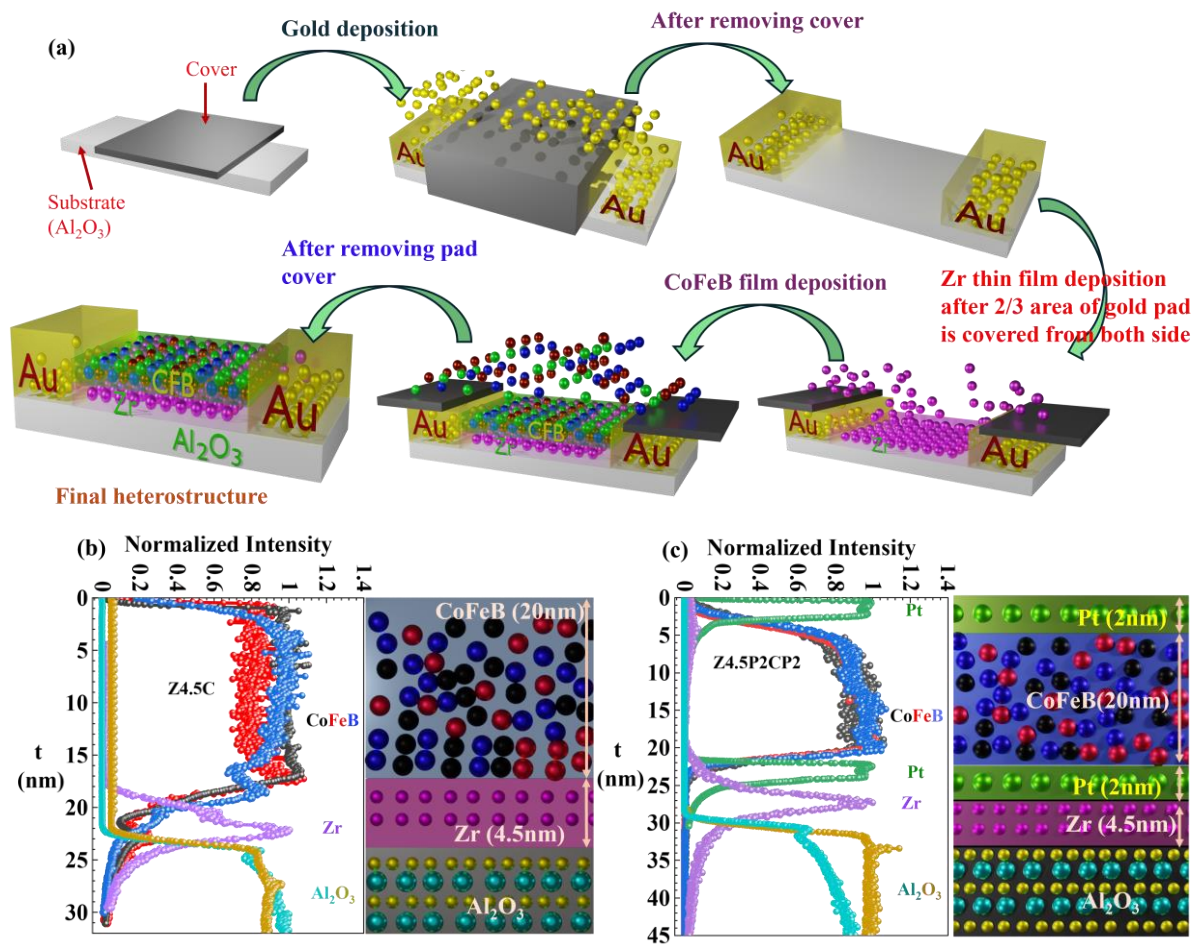

**Figure 1. Growth and characterization of CFB/Zr bilayer stacks**: (a) Schematic showing the sequence of steps undertaken in growing the heterostructure by sputtering with Au contacts pads required for electrical measurements. (b, c) Depth profile spectra (TOF-SIMS normalized intensity vs sputter time) of Z4.5C and Z4.5P2CP2 heterostructures.

## Results and Discussion

## Spin-orbital pumping ferromagnetic resonance measurements

Figure 2 illustrates the scheme of spin-orbit pumping ferromagnetic resonance (SOP-FMR) phenomenon. To initiate this phenomenon, a microwave-frequency signal having $H_{rf}$ magnetic field component is applied perpendicular to the magnetization (*M*) of the FM layer along with the external DC magnetic field *H*. Under the combined influence of DC magnetic field and rf-magnetic fields ($H_{rf}$), the magnetic moments start to precess around the equilibrium *M* and eventually reach to the resonance condition when frequency of $H_{rf}$ matches with the precession frequency of magnetic moment or called Lamor precession frequency of ferromagnet. At this ferromagnetic resonance condition, the spin angular momentum can transfer to the adjacent layer. If the SOC of the ferromagnet is significantly high, then spin (*S*) of the FM layer intertwines with its orbital angular momentum *L*, resulting in combined spin-orbital current ($J_{L,S}$). As the SOC strength of CFB is moderate and so it can generate combined spin and orbital current ($J_{L,S}$) via the intertwining of spin with its orbital angular moment. Under this resonance condition, $J_{L,S}$ can be pumped into the adjacent NM layer and converted into charge current through ISHE of spin current and IOHE of orbital current. The spin pumping parameter is defined by the enhancement of the Gilbert damping constant $\alpha_{eff}$. The detailed information about damping constant and FMR are described in the Supplementary section S1. Consequently, the efficiency of charge conversion from spin and orbital current transferred via FM/NM interface will depend on both the $\sigma_{SH}$ and $\sigma_{OH}$ of the NM materials. Whereas $\sigma_{SH}$ depends on the SOC of the NM layer[1], the $\sigma_{OH}$ is independent of SOC[22].

In the present work, Zr and CFB are chosen as the NM layer and FM layer, respectively. As the orbital Hall conductivity of Zr, $\sigma_{OH}^{Zr} = 5300 \frac{\hbar}{e} (\Omega - cm)^{-1}$ is much larger than its spin Hall conductivity, $\sigma_{SH}^{Zr} = -170 \frac{\hbar}{e} (\Omega - cm)^{-1}$ [20,24,25], it is expected that more orbital current from the CFB layer will be converted into charge current in the Zr layer compared to the spin current to charge current counterpart. Any amount of spin orbital current $J_{L,S}$ injected from CFB into the Zr is expected to get converted into the charge current via following three methods:

 (i) **ISHE**: Spin to charge current conversion attributed to ISHE ($J_C^{ISHE}$) is given by $\vec{J}_C^{ISHE} = \theta_{SH}^{Zr}(\vec{J}_S \times \vec{\sigma}_S)$, and is strongly dependent on the SOC strength of Zr which has a proportional relationship with the spin Hall angle of Zr ($\theta_{SH}^{Zr}$) and the spin polarization $\vec{\sigma}_S$ of the spin current propagating in Zr. Given the small and negative sign of $\sigma_{SH}$ of Zr, which is directly related to the $\theta_{SH}$, the spin to the charge current conversion is expected to be smaller [39–41]. As can be seen from the above-mentioned equation, the polarity of ISHE generated charge current is dependent on both the directions of spin current $\vec{J}_S$ and sign of $\sigma_{SH}$ [19].

**(ii) IOHE**: Orbital to charge current conversion via IOHE ($J_C^{IOHE}$) is given by $\vec{J}_C^{IOHE} = \theta_{OH}^{Zr}(\vec{J}_L \times \vec{\sigma}_L)$, where $\theta_{OH}^{Zr}$ is orbital Hall angle which is directly proportional to $\sigma_{OH}$, and $\vec{\sigma}_L$ is the polarization of orbital angular moment. The converted orbital to charge current (which is independent of the SOC strength), however, depends on the orbital texture present in the electronic states [25,39,42]. The spin current from CFB will be intertwining with this orbital texture of Zr and further gets converted into the charge current via IOHE. The polarity of $\vec{J}_C^{IOHE}$ will depend on the direction of orbital current ($\vec{J}_L$) and sign of ($\sigma_{OH}$). For the Zr, $\sigma_{OH}$ is opposite to the $\sigma_{SH}$ (i.e., $\vec{L}.\vec{S} < 0$). Therefore, the orbital charge current generated due to IOHE will be opposite to the polarity of the ISHE generated charge current [19,20].

**(iii) IOREE**: Orbital to charge current via IOREE, where $\vec{J}_C^{IOREE} = \lambda_{IOREE}(\hat{z} \times \delta\vec{L})$, where $\lambda_{IOREE}$ is the orbital to charge conversion efficiency from the two dimensional Rashba like states, and $\delta\vec{L}$ is the non-equilibrium orbital angular momentum current density caused by the transfer of orbital angular momentum in the Zr/CFB interface. The charge current polarity generated due to IOREE is independent of the orbital current direction or sign of $\sigma_{OH}^{Zr}$ [19,22].

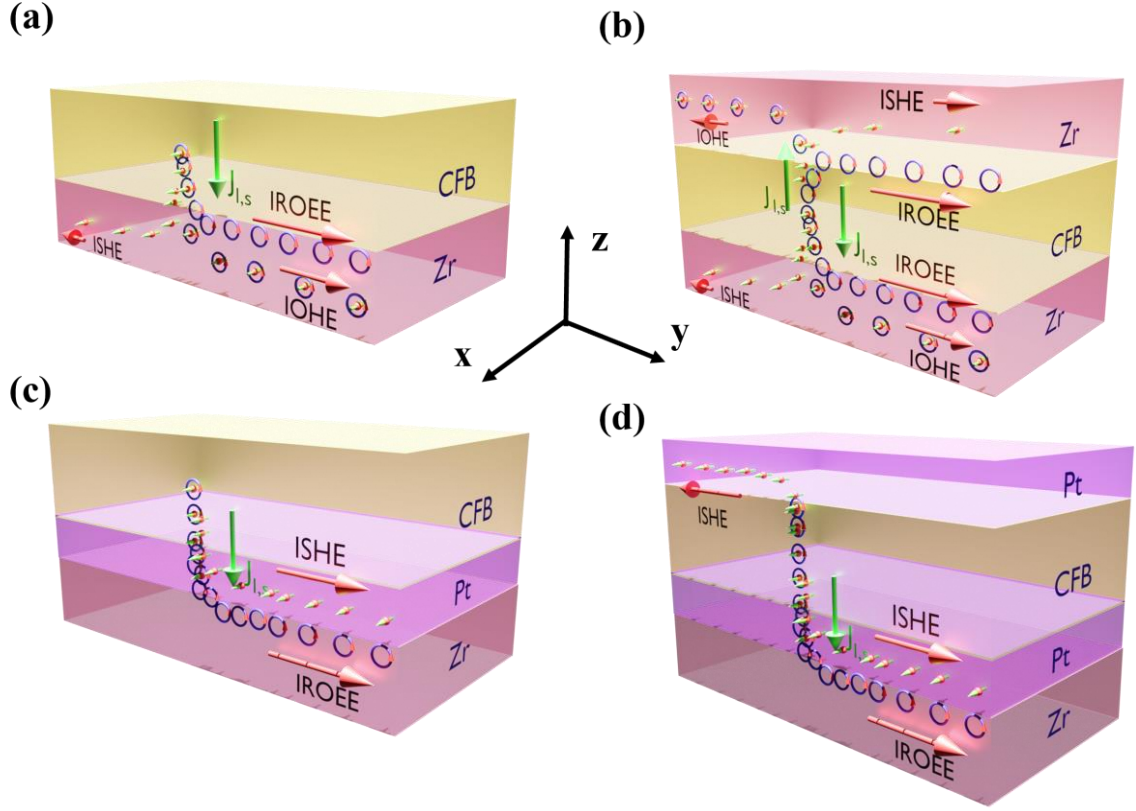

**Figure 2. Schematic Representations of Spin and Orbital Current Generation and their Charge Conversion in Zr/CFB Heterostructures: (a)** Schematic illustration of the spin and orbital pumping phenomenon in the Zr/CFB heterostructure. Spin current is generated in the ferromagnetic CFB layer, where the spin current in the bulk of CFB intertwines with the orbital angular momentum due to the moderate SOC strength of CFB. This process generates both spin and charge current simultaneously. These currents are pumped into the adjacent Zr layer, where the orbital current is converted into charge current via both the interfacial inverse orbital Rashba-Edelstein effect (IOREE) at the Zr/CFB interface and the inverse orbital Hall effect (IOHE) and inverse spin Hall effect (ISHE) within the bulk of Zr. **(b)** Schematic depiction of spin and orbital currents in the Zr/CFB/Zr structure. Spin and orbital currents are injected in both upward and downward directions into the Zr thin layers. Due to the ISHE, the spin current in the top Zr layer flows in the opposite direction compared to the bottom Zr layer. Similarly, driven by the IOHE, the orbital current in the bulk Zr flows in opposite directions for the top and bottom Zr layers. At the two Zr/CFB and CFB/Zr interfaces of CFB layer, the orbital currents get converted into charge current in the same direction via the IOREE effect [19,22]. **(c)** Schematic of spin-to-orbital conversion in Zr/Pt/CFB, where a Pt layer is inserted between the Zr and CFB layers. Here, the spin and orbital current generated by the spin-orbit pumping phenomena in CFB layer transferred into the adjacent Pt layer. The spin current directly converted into charge current due to inverse spin Hall effect in high SOC Pt layer, whereas remaining pure orbital current transfer into the adjacent Zr layer and convert into the charge current via IOREE and IOHE. **(d)** Schematic showing the cancellation of ISHE-induced voltage by the opposing polarity in a symmetric Pt/CFB/Pt structure. In this stack, the ISHE-generated spin currents from the top and bottom Pt layers cancel out each other, resulting in a net current that is purely orbital in nature, which eventually contribute to the charge current in Zr via IOREE and IOHE.

Figures 3(a-c) show the field dependence of the measured voltage signal $V_{DC}$ for the 6 GHz microwave frequency and 5dBm microwave power. The symmetric part of the underlying generated charge current is corresponding to the spin-to- and/or orbital-to-charge current(s)

while asymmetric part of generated signal arise due to rectification effects such as anisotropic magnetoresistance (AMR) and planar Hall effect (PHE). The $V_{DC}(H)$ is fitted using a Lorentzian function with symmetric ($V_{sym}$) and antisymmetric components ($V_{asym}$):

$$V_{DC} = V_{sym} \frac{\Delta H^2}{\Delta H^2 + (H - H_r)^2} + V_{asym} \frac{\Delta H (H - H_r)}{\Delta H^2 + (H - H_r)^2} \qquad (1)$$

here, $H$, $\Delta H$ and $H_r$ are applied magnetic field, linewidth, and resonance position of Lorentzian peak whereas.

The symmetric contribution stems from ISHE or IOREE signal. For current comparison, the symmetric voltage is divided by the total resistance $R$ of the sample to estimate the induced $I_{DC} = V_{sym}/R$. Figures 3(a-c) present the charge current generated by spin-orbital pumping for the Z4.5C, Z4.5P2C, and Z4.5P2CP2 stacks, respectively. In the Z4.5C stack, the charge current mainly arises from three contributions: ISHE, IOHE, and IOREE (assuming negligible ISREE), as described in the schematic in figure. 2a. As previously discussed, the $\sigma_{OH}$ is greater than the $\sigma_{SH}$, and the $\sigma_{SH}$ for the Zr is negative, whereas the $\sigma_{OH}$ is positive [20,25]. At the Zr/CFB interface, the inversion symmetry is broken due to the disparity of work function between CFB ($\varphi_{CFB} = 4.8\ eV$) and Zr ($\varphi_{CFB} = 4.05\ eV$), resulting in the formation of Rashba interface states[25,26,39].

As noted previously, under FMR condition, spin and orbital currents are generated simultaneously in the CFB and get transferred to the adjacent Zr layer. At the Zr/CFB interface, the orbital current is converted into charge current through the IOREE, while in the bulk of Zr, the spin angular moment is intertwined with the orbital angular momentum of the Zr and, together get converted into the charge current.[19] However, due to the negative $\sigma_{SH}$ of Zr, the spin current is converted into charge current through the ISHE in the opposite direction to that of the charge currents generated via IOREE and IOHE from the conversion of the orbital current. However, the ISHE signal is expected to be quite small due to weak SOC and low $\sigma_{SH}$ of Zr.

To determine whether the orbital-to-charge conversion is primarily due to IOREE or IOHE, we performed Zr thickness-dependent measurements ($t=2, 3, 4.5, 6,$ & $10$ nm). Figure 3d plots the charge current across different Zr thicknesses. The raw data of all the thickness dependent samples is presented in the Supplementary information S2. No significant variation in charge current magnitude was observed with Zr thickness up to 10 nm, thereby indicating that IOREE is the dominant contribution.

Now, to filter out the contribution of ISHE and IOHE, two additional engineered stacks, namely, CZ4.5C and Z4.5CZ4.5 were grown (See Table-1), making sure that spin and orbital currents from CFB layer must enter from the top and bottom surface of Zr in the designed experiment. In both IOHE and ISHE, the generated charge current depends on the direction of the orbital and spin currents and hence expected to cancel out when two combined spin-orbital currents travel to Zr with opposite direction (See supplementary S2). Figures 3d & 3e show that the charge current increases in the Z4.5CZ4.5 and CZ4.5C as compared to the Z4.5C samples. Compared to Z4.5C stack, the enhancement observed in case of Z4.5C is due to the additive effect of the charge currents generated at both the top and bottom interfaces (CFB/Zr and Zr/CFB and), having IOREE origin, as illustrated in the schematic of spin and orbital current in Figure. 2b.

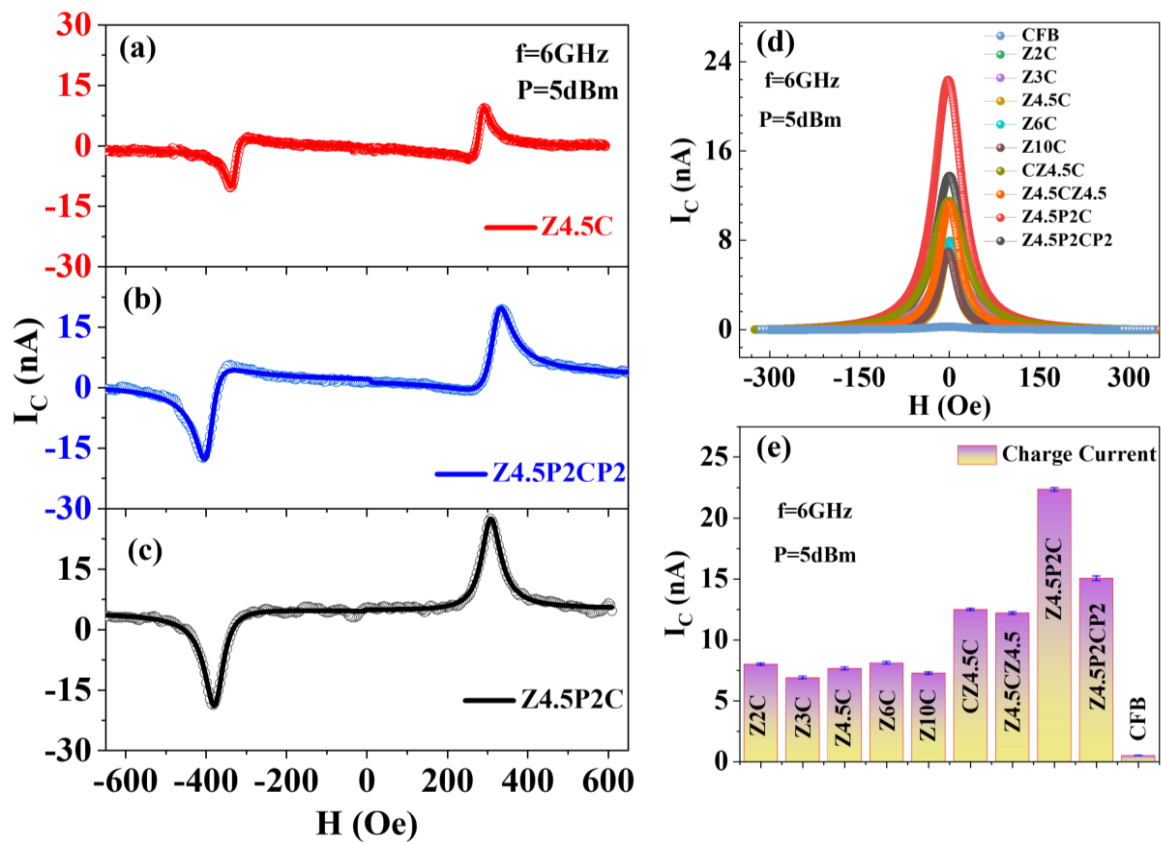

**Figure 3. Spin-Orbital Pumping and Charge Conversion Analysis in Zr/CFB and Zr/Pt/CFB stacks:** (a) Measured charge current in the ZC4.5 sample, generated through spin-orbital pumping, using a microstrip-based waveguide at an RF power of 5 dBm. (b) Measured charge current in the Z4.5P2CP2 sample, where opposing ISHE and IOHE contributions from the top and bottom Pt layers

are cancelled, isolating the charge current produced via the IOREE in the Zr layer. (c) Measured charge current in the Z4.5P2C sample, where the insertion of a Pt layer enhances the total charge current by generating pure orbital current through the high spin-orbit coupling of Pt. This orbital current is converted into charge current in the Zr layer via the IOREE effect. (d) Comparison of the symmetric part of the charge current across all samples, highlighting the individual contributions from ISHE, IOHE, and IOREE effects (e) Bar graph displaying the charge current contributions from spin and orbital currents across different samples. The charge current remains constant for the ZC2-ZC10 samples, indicating interface-dominated conversion, while a notable increase is observed when a Pt layer is inserted between Zr and CFB, due to the enhanced orbital contribution.

Next, a few more additional control experiments were further designed to inject a pure and higher amount of orbital current into the Zr layer. For this, a 2 nm Pt layer having small 1-1.5 nm spin diffusion length was inserted between the CFB and Zr layer (sample Z4.5P2C). Under FMR condition, both spin and orbital currents are transferred to the adjacent Pt layer. The spin current generated by CFB is transferred into the Pt layer via spin pumping, in which some fraction of the spin current gets converted into the charge current via ISHE and remaining fraction of the spin current intertwines with the orbital angular moment of Pt due to high SOC and generate a pure orbital current. The orbital current generated via Pt layer is further transferred into the Zr layer, where it is converted into charge current via IOREE as depicted in figure 2c. This results in an enhancement of total charge current, as shown in figure 3b. Hence, the total spin current has two contributions, one is from the ISHE in Pt layer and second is due to the IOREE in the Pt/Zr interface. So, the total charge current density ($J_C^{Z4.5P2C}$) in the Z4.5P2C stack can be expressed as[19]:

$$\vec{J}_C^{Z4.5P2C} = \theta_{SH}^{Pt}(\vec{J}_S \times \vec{\sigma}_S) + \lambda_{IOREE}^{Zr}(\hat{z} \times \delta\vec{L}) \qquad (2)$$

Where $\theta_{SH}^{Pt}$ is spin Hall angle of Pt, $\vec{J}_S$ is spin current density, $\vec{\sigma}_S$ is spin polarization, $\lambda_{IOREE}^{Zr}$ is inverse orbital Edelstein effect length of Zr, $\hat{z}$ is direction perpendicular to sample interface and $\delta\vec{L}$ is orbital surface current density [19,20,22].

To isolate the contribution of charge current generated purely by the orbital current, a sample is fabricated where the ISHE contribution from Pt cancels out *i.e.* putting Pt layers on both side of the CFB layer. The schematic of such heterostructure showing the spin and orbital current generation is shown in figure. 2d. This structure consists of Zr(4.5 nm)/Pt(2 nm)/CFB(20 nm)/Pt (2 nm) and is labelled as Z4.5P2CP2. In this case, the ISHE current density generated in the bottom Pt layer $\vec{J}_C^{Pt-Bottom}$ is cancelled out by the ISHE contributions from the top Pt layer $\vec{J}_C^{Pt-Top}$, leaving only the IOREE generated current density contribution in the Zr layer

$\vec{J}_C^{Zr}$ shown in figure 3c. The resulting current density $\vec{J}_C^{Z4.5P2CP2}$, generated purely by the orbital current via IOREE in Zr, can be expressed as:

$$\vec{J}_C^{Z4.5P2CP2} = \vec{J}_C^{Pt-Top} + \vec{J}_C^{Zr} + \vec{J}_C^{Pt-Bottom} \qquad (3)$$

where the top and bottom Pt layers, having same thickness, generate equal but opposite charge currents. Therefore, $\vec{J}_C^{Pt-Top} = -\vec{J}_C^{Pt-Bottom}$, and the net total current given by:

$$\vec{J}_C^{Z4.5P2CP2} = \vec{J}_C^{Zr} = \lambda_{IOREE}(\hat{z} \times \delta\vec{L}) \qquad (4)$$

From figures 3(d, e), where charge current measured for all the stacks listed in Table-1 is presented for quick comparison, one can clearly observe that the charge current in the Zr layer, among all the Z($t_{Zr}$)C type stacks, is dominated by the IOREE effect. It is independent of Zr layer thickness. The charge current in the sample Z4.5CZ4.5 and CZ4.5C is ~100% higher than the Z4.5C sample because of the spin orbit current in Zr injected from both top and bottom direction, which leads to enhancement in the charge current. While in the sample Z4.5P2C, the charge current enhanced ~280% of Z4.5C due to the additional charge current from Pt layer via ISHE and enhancement of orbital current in the Zr which leads to enhance orbital to charge current in Zr via IOREE. However, in the sample Z4.5P2CP2, the ISHE contributions get cancelled out and the resulting current is ~160% higher compared to that observed in case of the Z4.5C stack. The bar graphs in figure 3(e) demonstrate the comparison of charge current with thickness dependent and with and without Pt layer inserted between Zr and CFB (samples Z4.5P2C). The extracted values related to the parameters of spin current density, Gilbert damping constant, effective Hall angle and charge current densities are shown in the table 1 of supplementary file.

## Longitudinal Spin and Orbital Seebeck Effect (LSOSE) measurements

To further validate the results observed via spin-orbital pumping, longitudinal spin-orbital Seebeck effect (LSOSE) measurements were performed. Figure 4a provides a schematic illustration of the longitudinal measurement setup for the spin-orbital Seebeck effect. A temperature gradient (ΔT) was applied out-of-plane to the Zr(t)/CFB bilayer samples using a localized nichrome heater, while the generated voltage was detected with a Keithley 2182 nanovoltmeter. When an out of plane temperature gradient, $\nabla_z T$, is applied along the z-axis to the NM/FM system (here, Zr/CFB), along with in-plane applied external magnetic field ($H_x$), a spin current $J_S^{LSOSE} \parallel \nabla_Z T$ is generated due to thermally-induced magnonic wave along the z-axis in the FM layer. In the bulk of the CFB, this spin current intertwines with the orbital angular momentum due to the moderate SOC of CFB layer, thereby generating a combined spin orbital current ($J_S^{LSOSE}$). This spin orbital current is injected into the adjacent NM layer, where it is converted to charge current via IOREE at interface of Zr/CFB, and via IOHE and ISHE in the bulk of Zr. Additionally, an anomalous Nernst effect (ANE) contribution also arises due to the presence of the metallic FM layer[43]. To isolate the IOREE contribution, in the present study, thickness-dependent LOSEE measurements have been performed on the Zr/CFB heterostructures with Zr films thickness $t_{Zr}$=2, 3, 4.5, 6, 10 nm. Figure 4b, displays the total charge current ($I_{LSOSE+ANE}$) measured along the sample plane for the representative Z4.5C heterostructure. The magnetic field was swept from negative to positive in presence of the different values of out of plane temperature gradient (ΔT=5, 10, 17, 26K). The field required to saturate the charge remained constant irrespective of the value of the temperature gradient. Thickness dependent $I_{LSOSE+ANE}$ hysteresis curve for samples Z2C, Z3C, Z4.5C, Z6C, and Z10C are presented in supplementary file S3.

To verify that thermally-induced magnonic wave driven IOREE effect is predominantly involved in injecting the spin-orbital current into Zr from the top and bottom directions in CZC type stacks, similar measurements were also performed over the sample CZ4.5C. Measurements conducted on the CZ4.5C sample demonstrated an enhanced charge current, as detailed in the supplementary file S3, confirming that the observed current arises primarily from the Rashba effect and contribution from both the interfaces gets added due to the direction independent nature of the IOREE phenomena. Moreover, to enhance the charge current ($I_{LSOSE+ANE}$) induced in Zr via orbital to charge conversion, a 2nm Pt layer was inserted between the Zr and CFB in Z4.5P2C sample. The thermally generated spin current from the CFB layer due to magnonic wave diffused into the Pt layer, where it is converted partly into the orbital current via intertwining of spin and orbital moments presence in the Pt layer, and partly into the charge current via ISHE within Pt. Meanwhile, the pure orbital current $J_{l,s,}$ was transferred into the Zr layer, where it is further converted into the charge current via IOREE effect in the Zr film. The resulting charge current hysteresis loops are shown in figure 4c for different magnitude of temperature gradient ($\nabla_Z T$). Notably, the measured charge current ($I_{LSOSE+ANE}$) in Z4.5P2C increased nearly 300% compared to the Z4.5C sample (see figure. 4c). This is because it also contains the ISHE contribution from Pt , along with the enhanced IOREE contribution from the Pt/Zr interface. Finally, to eliminate the ISHE contribution from Pt, an additional Pt layer of the same thickness deposited on the top of CFB in Z4.5P2CP2 sample. Since, ISHE current is dependent on spin current direction, the charge induced currents from the top and bottom Pt layer cancelled each other, leaving the resulting current to arise purely from orbital-to-charge conversion in Zr layer and ANE in CFB layer. Figure 4d displays this resultant current, dominated by IOREE, while the ANE contribution is relatively small, as discussed in the supplementary file. This also proves the claim of spin-orbital intertwining in Pt that results enhanced spin-orbital current that traversed to the Zr by comparing the charge current output from Z4.5C and Z4.5P2CP2. The slight enhancement in the charge current in the Z4.5P2CP2 compared to Z4.5C is indeed an experimental proof of the spin-orbital intertwining in Pt that enhances the spin-orbital current to the Zr. The LOSEE signal, as shown in figure 4e, exhibits the linear relationship with ΔT, indicating that the ANE charge contribution from CFB is minimal, while the charge current from Zr layer, particularly in presence of a Pt layer, significantly increases. A bar graph in figure 4f compares the charge contributions at ΔT = 10 K, clearly demonstrating that the orbital-induced charge current in Zr is higher than the ISHE-induced charge current in Pt.

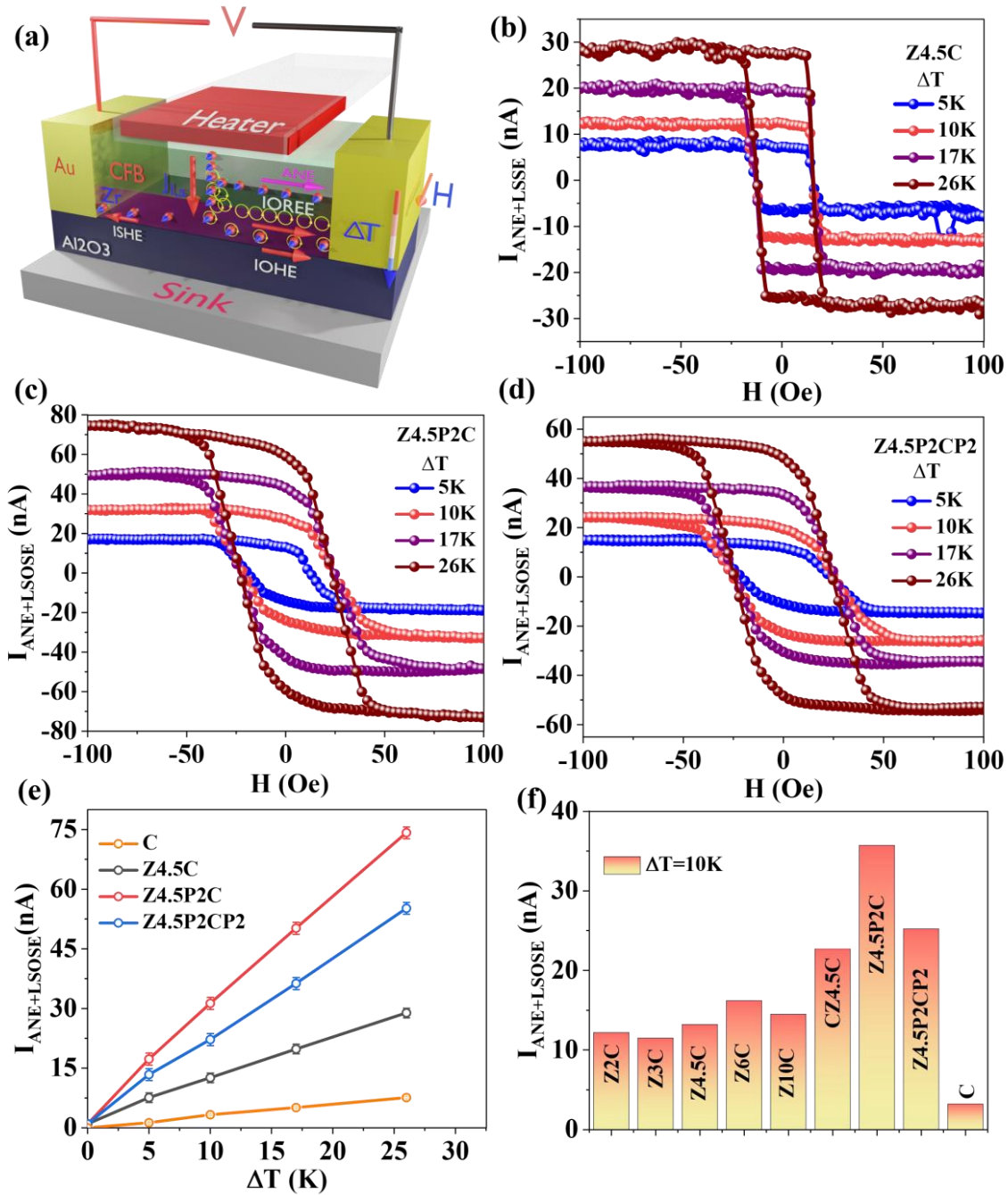

**Figure 4. Longitudinal Spin-Orbital Seebeck Effect (LSOSE) and Charge Current response and their comparative analysis in Zr/CFB and Zr/Pt/CFB heterostructures:** (a) Schematic illustration of the measurement setup for the spin Seebeck effect, where an out-of-plane temperature gradient (ΔT) is applied across the Zr/CFB heterostructures. Total charge current ($I_{LSOSE+ANE}$) as a function of the applied magnetic field for the samples is shown in (b) Z4.5C, (c) Z4.5P2C, and (d) Z4.5P2CP2, measured at various temperature gradients (ΔT = 5, 10, 17, 26 K). The voltage/current was measured in the saturation regime of CFB. (e) Comparison of total charge current ($I_{LSOSE+ANE}$) as a function of temperature gradient across different samples at saturation magnetization conditions, indicating the contributions from both spin and orbital effects. (f) Comparison of total generated current ($I_{LSOSE+ANE}$) at ΔT = 10 K, for all the measured samples, showcasing the enhancement of the orbital current contribution due to the presence of Pt.

## Discussion

The field of spintronics has long been dominated by the generation, propagation, and manipulation of electron spin, driven by key breakthroughs such as the spin Hall Effect and the Rashba effect. While spin has received significant attention, the orbital angular momentum—an equally fundamental property of electrons—has so far largely overlooked. Traditionally, the orbital degree of freedom was seen as merely a contributor to spin-orbit coupling, rather than as an independent entity. It was only with the recent discovery of the orbital Hall effect that the importance of orbital currents began to be recognized. Unlike spin, which is often constrained by material properties and strong spin-orbit coupling, orbital currents can be generated in a wider range of materials, including those with weaker spin-orbit interactions. In this study, we sought to harness the electron's orbital degree of freedom alongside the spin degree of freedom using a light element like Zr, combined with an ultrathin Pt layer and a ferromagnetic CFB layer. The efficiency of spin-orbital current to charge current conversion ($\theta_{eff}$) was estimated using $\theta_{eff} = J_{LS}/J_C$, where $J_{LS}$ was calculated from the difference of Gilbert damping constant of FM/NM and FM layers (detailed in supplementary file S1, with parameters listed in Table S1). The charge current density $J_C$ was calculated using the expression $J_C = I_C/(w \times t_{NM})$, where $w$ is the width of sample and $t_{NM}$ is the thickness of NM layer. Figure 5 presents a comparison of $\theta_{eff}$ for Zr based heterostructures (present work) with various systems reported in the literature that primarily focus on heavy metals (HM) and topological insulators (TI), which only exploit the spin degree of freedom while negligence of the orbital component[11,44–53]. Notably, the spin-orbital current to charge current conversion efficiency in Zr/Pt/CFB heterostructures is significantly higher than in conventional heterostructures that rely solely on spin-to-orbital to charge conversion as compared in figure 5.

Our work opens new avenues for advancing spintronic technologies by leveraging both spin and orbital angular momentum, potentially enabling more versatile, efficient, and powerful devices. The realization that the orbital degree of freedom can be manipulated just as effectively as spin represents a paradigm shift in spintronics, expanding the toolkit available for developing next-generation electronic devices.

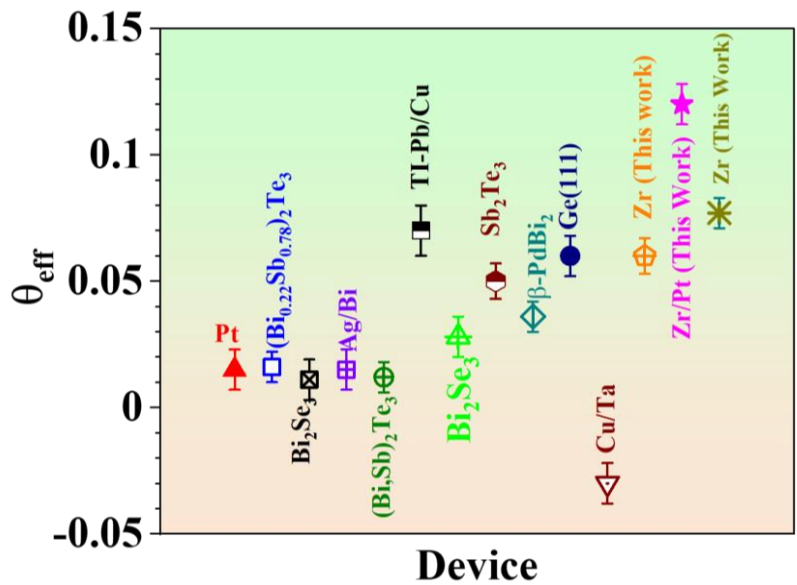

**Figure 5. Comparison of effective Spin-Orbital Hall angles across various heterostructures:** Comparison of effective spin orbital Hall angle ($\theta_{eff}$) for Zr, Zr/Pt, and Zr/Pt/FM/Pt heterostructures with values reported in the literature for HM and TI systems [11,44–53]. The effective Hall angle includes both spin and orbital contributions in the Zr and Zr/Pt systems.

# Conclusion

In summary, two physically distinct methodologies namely Spin-Orbital Pumping (SOP) via Ferromagnetic Resonance (FMR) and the Longitudinal Spin-Orbital Seebeck Effect (LSOSE) techniques were utilized to explore the generation, propagation, manipulation and conversion of spin-orbital currents in Zr/CFB and Zr/Pt/CFB heterostructures. It was convincingly found that, in the Zr/CFB heterostructure the primary contributor to the spin-orbital to charge current conversion is the interfacial IOREE phenomena, as it is found to be consistent within the range of Zr thickness. In addition, by incorporating Pt layer, which has strong SOC, of thickness greater than spin diffusion length (~1–1.5 nm), spin-orbital current density is enhanced that results in threefold enhancement in the final converted charge current output. Whereas the spin current gets converted to charge current via the ISHE in Pt, the combined spin-orbital current gets transferred into Zr wherein it gets converted into charge current via the IOREE. The insertion of a Pt layer enables to utilize both the spin current and the enhanced spin-orbital current conversion. Further, to eliminate ISHE contributions from Pt, and to solely estimate the IOREE contribution, similar Pt layer was deposited on both the top and bottom sides of the CFB layer. This configuration enabled us to cancel out the ISHE effects. The final converted charge current output was still found to be double that of the reference Zr/CFB heterostructure. This is possibly because of the enhanced orbital-to-charge conversion in Zr due to the spin-orbital intertwining in Pt. These findings represent the first experimental report of such a high orbital-induced charge current in Zr thin films, as verified by both SOP-FMR and LSOSE techniques, illustrating the potential for utilizing Zr in spintronic applications.

# Acknowledgment

N. Kumar acknowledges the Council of Scientific & Industrial Research (CSIR), Government of India, for financial assistance. We acknowledge the Department of Physics at IIT Delhi for the XRD facility for XRR measurements and the Central Research Facility of IIT Delhi for TOF-SIMS for depth profiling measurements.